\documentclass[pra,twocolumn,superscriptaddress,showpacs]{revtex4}
\usepackage{graphics} 
\usepackage{epsfig} 
\usepackage{mathptmx} 
\usepackage{times} 
\usepackage{amsthm}

\usepackage{amssymb, amsmath, mathrsfs}
\usepackage{epstopdf} 
\usepackage{thmtools}
\usepackage{subeqnarray}

\DeclareMathOperator{\rank}{rank} %
\DeclareMathOperator{\diag}{diag} %
\DeclareMathOperator{\ran}{range} %
\DeclareMathOperator{\tr}{tr} %
\usepackage{calc,pifont} 
\theoremstyle{definition}
\newtheorem{thm}{Theorem}

\allowdisplaybreaks 
\usepackage{xcolor,cancel}

\begin{document}
 \title{Preparation of bipartite bound entangled Gaussian states in quantum optics}
\author{Shan Ma}
\affiliation{School of Automation, Central South University, Changsha 410083, China}
\affiliation{State Key Laboratory of Quantum Optics and Quantum Optics Devices,\\ Shanxi University, Taiyuan 030006, China}
\author{Matthew J. Woolley} 
\affiliation{School of Engineering and Information Technology, UNSW Canberra, ACT 2600, Australia}
\author{Xiaojun Jia} 
\affiliation{State Key Laboratory of Quantum Optics and Quantum Optics Devices,\\ Shanxi University, Taiyuan 030006, China}
\author{Jing Zhang} 
\email{jzhang74@yahoo.com, jzhang74@sxu.edu.cn}
\affiliation{State Key Laboratory of Quantum Optics and Quantum Optics Devices,\\ Shanxi University, Taiyuan 030006, China}

\begin{abstract}
The positivity of the partial transpose  is  in general only a necessary condition for separability. There exist quantum states that are not separable, but nevertheless are positive under partial transpose. States of this type are known as  bound entangled states meaning that these states are entangled but they do not allow distillation of pure entanglement by means of local operations and classical communication (LOCC).  We present a parametrization of a class of $2\times 2$ bound entangled Gaussian states for bipartite continuous-variable quantum systems with two modes on each side. We propose an experimental protocol for preparing a particular bound entangled state in quantum optics. We then discuss the robustness properties of this protocol with respect to the occupation
number of  thermal inputs and the degrees of squeezing.
\end{abstract}  


\maketitle

Entanglement is a striking property of quantum mechanics, being central in most quantum information technologies. One of  the most fundamental problems in quantum information is to determine whether a quantum state is entangled or not.  During the last three decades, considerable effort has been devoted to solving this problem. Though it has
not yet been completely solved, a great deal of progress has been made. 
A critical progress is the development of an elegant criterion, known as partial transpose, for studying separability~\cite{HHH96:pla,P96:prl}. The partial transpose corresponds physically to a local 
time reversal operation \cite{HHH98:prl}. The positivity of the partial transpose  provides a necessary condition for separability. 
In some restricted cases, this criterion turns out to be  also  sufficient. To be  specific, for discrete-variable quantum systems, the positive partial transpose (PPT) criterion is necessary and sufficient for separability of $2\times 2$ and $2\times 3$ dimensional systems~\cite{HHH96:pla,P96:prl}. However, for higher dimenional systems (e.g., $3\times 3$ and $2 \times 4$ dimensional cases), this criterion fails to be sufficient for separability~\cite{P97:pla}.  For continuous-variable quantum systems, the PPT criterion is necessary and sufficient for separability of continuous-varible systems of $1\times n$ oscillators in a joint Gaussian state~\cite{WW01:prl,S00:prl,DGCZ00:prl}. However, for higher dimensions (e.g., a  continuous-variable system of $2 \times 2$ oscillators in a Gaussian state), this criterion fails to be sufficient for separability~\cite{WW01:prl}. For some very special classes of $n\times m$ Gaussian states (e.g., bisymmetric Gaussian states~\cite{SAI05:pra}, isotropic Gaussian states~\cite{HW01:pra,GECP03:QIC,BR03:pra,LSA18:njp}), the PPT criterion is also necessary and sufficient for separability. 

The PPT criterion is in general not a sufficient condition for separability.  There exist non-separable (entangled) states with positive partial transpose. These states are known as bound entangled states meaning that these states are entangled, but their entanglement cannot be distilled into maximally entangled states via LOCC~\cite{HHH98:prl}. 
Examples of bound entangled states have been found for discrete-variable quantum systems~\cite{P97:pla,S01:pra}, as well as for continuous-variable quantum systems~\cite{WW01:prl,HL00:prl}.  A large effort
has gone into the analysis and detection of bound entanglement~\cite{SST03:prl,ZNZG10:pra,Z11:pra,DSHPES11:prl,JZWZXP12:prl,SDSV14:pra,SES16:pra,SGSSK18:quantum,SP18:pra}, as well as their applications in steering~\cite{MGHG14:prl},  metrology~\cite{TV18:prl}, entanglement
activation \cite{HHH99:prl,M06:prl}, quantum key distribution~\cite{HHHO05:prl,AH09:pra}  and nonlocality~\cite{VB14:nc}. In particular, it has been shown recently that there exist bound entangled states  that can  be used for steering~\cite{MGHG14:prl}  and that can even violate a Bell inequality~\cite{VB14:nc} for discrete-variable systems. These results disprove a longstanding conjecture known as the Peres conjecture which states that bound entangled states cannot violate any Bell inequality~\cite{P99:fp,P13:pra}. For the continuous-variable case, it has been shown that bound entangled Gaussian states cannot display steering under Gaussian measurements~\cite{KLRA15:prl,JKN15:jpa}. However, it is still an open question whether there exist bound entangled Gaussian states that can violate a Bell inequality and thus can display EPR steering when non-Gaussian measurements are involved. 

We consider continuous-variable entanglement with   Gaussian states which serve as key resources for Gaussian quantum information processing~\cite{BP03:book,weedbrook12:rmp}. Gaussian states arise naturally in quantum optics and are completely characterized by the first and second moments of canonical operators. The first moments (i.e., mean) contain no information about  entanglement and can be shifted to zero by local unitaries, thus are irrelevant for our purpose. All the information
about  entanglement of Gaussian states is contained in
the second moments (i.e., covariance matrix). In this work, we  parametrize a class of $2\times 2$ bound entangled Gaussian states by characterizing their covariance matrices. Our parametrization result provides a simple and accurate way to obtain examples of bound entangled states. It does not rely on numerical computation that may be significantly affected by rounding errors. In addition, for a particular bound entangled state, we propose an experimental protocol for preparing 
it in quantum optics.  We also investigate the robustness properties of this protocol with respect to the occupation
number of  thermal inputs and the parameters of squeezing components. Though we only consider the preparation of one particular state in this work, the method we use here can be applied to any other bound entangled Gaussian state to obtain a corresponding preparation scheme. 

The experimental preparation and verification of continuous-variable bound entanglement has been conducted in quantum  optics~\cite{DSHPES11:prl}. It is known that continuous-variable bound entanglement is a rare phenomenon \cite{HHHH09:rmp}. It is  possible that a bound entangled state has both free entangled and separable states very nearby. The experimental preparation of bound entanglement generally requires a high-precision control over the system parameters and hence is difficult to implement. On the other hand, the verification of bound entanglement in the laboratory is also a challenging task since the entanglement and PPT tests are sometimes very sensitive to experimental errors and certification requires a very careful analysis of the experimental data. Different from the generating scheme introduced in Ref.~\cite{DSHPES11:prl}, we study the preparation of bound entanglement using an analytical method. We first perform some decompositions (specifically, thermal decomposition of a covariance matrix~\cite{W36:ajm} and Euler decomposition of a canonical unitary~\cite{ADMS95:Pramana,B05:pra}) on a particular bound entangled state. These decompositions are then translated into an optical network with input fields such  that the target bound entangled state is generated. All of these procedures are completed analytically. This purely
analytical treatment allows us to have
a more precise understanding of how bound entanglement is generated in quantum optics. In addition, the scheme presented in this work allows us to further investigate the robustness of bound entanglement preparation by varying some parameters in the optical system and pinpoint a region in the parameter space such that bound entanglement can exist.   
 
Let us consider a bosonic system of $n$ modes. Each mode is characterized by a pair of quadrature field operators $\{\hat{q}_{k},\hat{p}_{k}\}_{k=1}^{n}$ (position and momentum operators). We arrange the operators $\hat{q}_{k}$ and $\hat{p}_{k}$ in a vector of operators $\hat{X}=(\hat{q}_{1},\hat{p}_{1},\cdots,\hat{q}_{n},\hat{p}_{n})^{T}$. The canonical commutation relations for $\hat{X}_{k}$ take the form (with $\hbar=1$)
\begin{align}
[\hat{X}_{j},\;\hat{X}_{k}]=i\sigma_{jk},
\end{align}
where $\sigma_{jk}$ is the generic entry of the $2n\times 2n$ symplectic matrix $\sigma=\bigoplus\limits_{k=1}^{n}\begin{pmatrix}
0 &1\\
-1 &0
\end{pmatrix}$. 
 We introduce the Weyl displacement operator $W_{\xi}=exp(i\hat{X}^{T}\sigma\xi )$ with $\xi \in \mathbb{R}^{2n}$. Then an arbitrary continuous-variable quantum state $\rho$ can be fully described  in terms of its symmetrically ordered characteristic function $\chi$ defined by $\chi(\xi)=\tr(\rho W_{\xi})$. Gaussian states are bosonic states with a Gaussian characteristic function. Gaussian states are completely characterized by the first two moments of the canonical operators $\hat{X}_{k}$. The first moment is called the mean value which is defined as   the vector $\bar{X}:=\langle \hat{X} \rangle$ with $\bar{X}_{k}=\tr(\rho \hat{X}_{k})$. The second moment is called the covariance matrix $\gamma$ whose arbitrary element is defined by  $\gamma_{jk}=\langle \triangle\hat{X}_{j}\triangle\hat{X}_{k}+\triangle\hat{X}_{k}\triangle\hat{X}_{j} \rangle$, where $\triangle\hat{X}_{j}:=\hat{X}_{j}-\langle \hat{X}_{j}\rangle$.   
The covariance matrix $\gamma$ is a $2n\times 2n$ real and symmetric matrix which must satisfy the uncertainty principle~\cite{SMD94:pra}
\begin{align}
\gamma+i\sigma\ge 0. \label{covariance}
\end{align} This matrix inequality is also a sufficient condition for a real, symmetric matrix $\gamma$ to represent a covariance matrix of a Gaussian state. That is, for every real symmetric matrix $\gamma$ satisfying the inequality   \eqref{covariance}, there exists a Gaussian state with this $\gamma$ as its covariance matrix. The matrix inequality \eqref{covariance} implies $\gamma>0$. 

Suppose we have two bosonic systems $A$ with $n$ modes and $B$ with $m$ modes and the quantum state of the global bipartite  system $A+B$ is in a Gaussian state. By definition, a  quantum state of a bipartite system is   separable if its total density operator can be expressed as a convex sum of product states $\rho=\sum_{k}\eta_{k}\rho_{k,A}\otimes\rho_{k,B}$ where $\eta_{k}\ge 0$ and $\sum_{k}\eta_{k}=1$ \cite{W89:pra}. A state is called entangled if it is not separable. For Gaussian states, all of the entanglement properties are  contained in the covariance matrices $\gamma$. It was shown in Ref.~\cite{WW01:prl} that  
a Gaussian state is separable if and only if there exist two real symmetric matrices $\gamma_{A}\ge i\sigma_{A}$ and $\gamma_{B} \ge i\sigma_{B}$ such that
\begin{align} 
\gamma\ge \gamma_{A}\oplus \gamma_{B}. \label{separable}
\end{align}
The necessary and sufficient condition \eqref{separable} can be further simplified as $\gamma\ge \gamma_{A}\oplus i \sigma_{B}$ \cite{LSA18:njp}. Although the condition \eqref{separable} is very useful in demonstrating that some particular quantum states are entangled~\cite{WW01:prl,GKLC01:prl, HE06:njp}, it cannot be directly applied to an arbitrary state, since the analytical determination of  $\gamma_{A,B}$ is in general not possible.

In Ref.~\cite{WW01:prl}, it was also shown that a Gaussian state has positive partial transpose if and only if 
\begin{align}
\gamma+i\tilde{\sigma}\ge 0, \label{PPT}
\end{align} where $\tilde{\sigma}=(-\sigma_{A})\bigoplus\sigma_{B}
$. Recall that the PPT criterion provides a necessary condition for  separability. If a Gaussian state is separable, then it must have a PPT covariance matrix $\gamma$  satisfying \eqref{PPT}. However, the converse is in general not true. There exist non-separable Gaussian states with a PPT covariance matrix. This type of Gaussian state is known as a bound entangled Gaussian state. For continuous-variable quantum systems, the $2\times 2$ case is the simplest case in which bound entanglement exists.  A particular example of a $2\times 2$ bound entangled Gaussian state can be found in Ref.~\cite{WW01:prl}. Now we attempt to generalize this example, and provide a parametrization of a class of $2\times 2$ bound entangled Gaussian states. 

First combining the conditions \eqref{covariance} and  \eqref{PPT}, we see that a real symmetric matrix $\gamma$ is a PPT covariance matrix if and only if $\gamma+i\sigma\ge 0$ and $\gamma+i\tilde{\sigma}\ge 0$. It is a minimal PPT covariance matrix if any PPT covariance matrix $\gamma'$ satisfying $ \gamma' \le \gamma$ must be equal to $\gamma$~\cite{WW01:prl}.  A minimal PPT covariance matrix $\gamma$ is separable if and only if it is a direct sum, i.e.,   $\gamma= \gamma_{A}\oplus \gamma_{B}$ where $\gamma_{A,B}$ correspond to pure states. A PPT covariance matrix $\gamma$ is minimal if and only if $\gamma+i\sigma$ and $\gamma+i\tilde{\sigma}$ do not majorize a common nonzero real symmetric positive semidefinite matrix; that is, there is no \emph{real} vector $\zeta\ne 0$ that is in the support of both matrices: $\gamma+i\sigma$ and $\gamma+i\tilde{\sigma}$. This is the case if and only if the PPT covariance matrix $\gamma$ satisfies $\ran(\gamma+\sigma\gamma^{-1}\sigma)\cap \ran(\gamma+\tilde{\sigma}\gamma^{-1}\tilde{\sigma})=\{0\}$.

Motivated by the bound entangled state example proposed in Ref.~\cite{WW01:prl}, we consider a covariance matrix of the form
\begin{align}
\gamma=\begin{pmatrix}
\gamma_{11} &0 &0 &0 &\gamma_{15} &0 &0 &0\\
 0          &\gamma_{22} &0 &0 &0&0 &0  &\gamma_{28}\\
 0          &0     &\gamma_{33} &0 &0 &0 &\gamma_{37} &0\\
 0 &0 &0 &\gamma_{44} &0 &\gamma_{46} &0 &0\\
 \gamma_{15} &0&0 &0 &\gamma_{55} &0 &0 &0 \\
 0 &0 &0 &\gamma_{46} &0 &\gamma_{66} &0 &0\\
 0 &0 &\gamma_{37} &0 &0 &0 &\gamma_{77} &0\\
 0 &\gamma_{28} &0 &0 &0 &0 &0 &\gamma_{88}
\end{pmatrix}. \label{r}
\end{align}
This matrix $\gamma$ has a relatively simple form with no correlations between position and momentum operators, but still it can exhibit bound entanglement as we will show. A direct calculation shows that any matrix $\gamma$ of the form~\eqref{r}   commutes with a diagonal matrix $S$  with diagonal elements $(1,1,-1,-1,1,-1,-1,1)$. That is, $S\gamma=\gamma S$. Also,  $S\sigma=-\tilde{\sigma}S$. Thus $\gamma+i\sigma$ and $\gamma-i\tilde{\sigma}=S(\gamma+i\sigma)S$ are unitarily similar. In this case, $\gamma$ is a PPT covariance matrix if and only if $\gamma+i\sigma \ge 0$, which is equivalent to the positive semidefiniteness of $\begin{pmatrix}
\gamma &\sigma\\
\sigma^{\top} &\gamma
\end{pmatrix}\ge 0$. By Schur complement, this is further equivalent to $\gamma>0$ and $\gamma+\sigma\gamma^{-1}\sigma\ge 0$. To simplify analysis,  it is useful to introduce a permutation matrix:  
\begin{align*}
P=\begin{pmatrix}
     1     &0     &0     &0     &0     &0     &0     &0\\
     0     &0     &1     &0     &0     &0     &0     &0\\
     0     &0     &0     &0     &1     &0     &0     &0\\
     0     &0     &0     &0     &0     &0     &1     &0\\
     0     &1     &0     &0     &0     &0     &0     &0\\
     0     &0     &0     &1     &0     &0     &0     &0\\
     0     &0     &0     &0     &0     &1     &0     &0\\
     0     &0     &0     &0     &0     &0     &0     &1
\end{pmatrix}. 
\end{align*}
Using $P$ as a permutation matrix, we find that  $\sigma':=P\sigma P^{T}=\begin{pmatrix}
0 &I\\
-I &0
\end{pmatrix}$  and $\gamma':=P\gamma P^{T}=\gamma_{1}'\oplus\gamma_{2}'$  where $\gamma_{1}'=\begin{pmatrix}
\gamma_{11} &0 &\gamma_{15} &0\\
0 &\gamma_{33} &0 &\gamma_{37}\\
\gamma_{15} &0 &\gamma_{55} &0\\
0 &\gamma_{37} &0 &\gamma_{77}
\end{pmatrix}$ and $\gamma_{2}'=\begin{pmatrix}
\gamma_{22} &0 &0 &\gamma_{28}\\
0 &\gamma_{44} &\gamma_{46} &0\\
0 &\gamma_{46} &\gamma_{66} &0\\
\gamma_{28} &0 &0 &\gamma_{88}
\end{pmatrix}$. Thus, it suffices to check $\gamma'>0$ and $\gamma'+\sigma'\gamma'^{-1}\sigma'\ge 0$. This happens if and only if $\gamma_{2}'>0$ and $\gamma_{1}'-\gamma_{2}'^{-1}\ge 0$. 

On the other hand, we want $\gamma$ to be a minimal PPT covariance matrix. This happens if $\ran(\gamma+\sigma\gamma^{-1}\sigma)\cap \ran(\gamma+\tilde{\sigma}\gamma^{-1}\tilde{\sigma})=\{0\}$. Further analysis shows that $\gamma$ is a minimal PPT covariance matrix if $\rank(\gamma+\sigma\gamma^{-1}\sigma, \gamma+\tilde{\sigma}\gamma^{-1}\tilde{\sigma})=\rank(\gamma+\sigma\gamma^{-1}\sigma)+\rank(\gamma+\tilde{\sigma}\gamma^{-1}\tilde{\sigma})$~\cite{MS74:lma}. 
In order for $\gamma$ to correspond to an entangled state,  we assume $\gamma_{15}\ne 0$, $\gamma_{28}\ne 0$, $\gamma_{37}\ne 0$, and $\gamma_{46}\ne 0$ such that $\gamma$ is of a nonblock diagonal form. In this case, it can be shown that $\rank(\gamma+\sigma\gamma^{-1}\sigma, \gamma+\tilde{\sigma}\gamma^{-1}\tilde{\sigma})=8$. Since  $\rank(\gamma+\sigma\gamma^{-1}\sigma)=\rank(\begin{pmatrix}
\gamma &\sigma\\
\sigma^{\top} &\gamma
\end{pmatrix})-8=2\rank(\gamma+i\sigma)-8=2\rank(\gamma+i\tilde{\sigma})-8=\rank(\begin{pmatrix}
\gamma &\tilde{\sigma}\\
\tilde{\sigma}^{\top} &\gamma
\end{pmatrix})-8=\rank(\gamma+\tilde{\sigma}\gamma^{-1}\tilde{\sigma})$, we obtain $\rank(\gamma+\sigma\gamma^{-1}\sigma)=\rank(\gamma+\tilde{\sigma}\gamma^{-1}\tilde{\sigma})=4$. It follows that $\rank(\gamma'+\sigma'\gamma'^{-1}\sigma')=4$; that is $\rank(\gamma_{1}'-\gamma_{2}'^{-1})+\rank(\gamma_{2}'-\gamma_{1}'^{-1})=4$. Since $\gamma_{2}'-\gamma_{1}'^{-1}=\gamma_{2}'(\gamma_{1}'-\gamma_{2}'^{-1})\gamma_{1}'^{-1}$, we have $\rank(\gamma_{2}'-\gamma_{1}'^{-1})=\rank(\gamma_{1}'-\gamma_{2}'^{-1})=2$.
Since $\gamma_{1}'-\gamma_{2}'^{-1}\ge 0$, we take  
\begin{align*}
&\gamma_{1}'-\gamma_{2}'^{-1}\\
=&\begin{pmatrix}
\gamma_{11} &0 &\gamma_{15} &0\\
0 &\gamma_{33} &0 &\gamma_{37}\\
\gamma_{15} &0 &\gamma_{55} &0\\
0 &\gamma_{37} &0 &\gamma_{77}
\end{pmatrix}-
\begin{pmatrix}
d_{11} &0 &0 &d_{14}\\
0 &d_{22} &d_{23} &0\\
0 &d_{23} &d_{33} &0\\
d_{14} &0 &0 &d_{44}
\end{pmatrix}\\
=&\begin{pmatrix}
\beta_{1}\alpha_{1} &\alpha_{1}\\
\alpha_{2} &-\beta_{1}\alpha_{2}\\
\beta_{2}\alpha_{3} &\alpha_{3}\\
\alpha_{4} &-\beta_{2}\alpha_{4}
\end{pmatrix}\begin{pmatrix}
\beta_{1}\alpha_{1} &\alpha_{1}\\
\alpha_{2} &-\beta_{1}\alpha_{2}\\
\beta_{2}\alpha_{3} &\alpha_{3}\\
\alpha_{4} &-\beta_{2}\alpha_{4}
\end{pmatrix}^{T}.
\end{align*}
We take $d_{14}=(\beta_{2}-\beta_{1})\alpha_{1} \alpha_{4}$, $
d_{23}=(\beta_{1}-\beta_{2})\alpha_{2} \alpha_{3}$,
$d_{11}=\alpha_{5}$,
$d_{22}=\alpha_{6}$,
$d_{33}=(\beta_{1}-\beta_{2})^{2}\alpha_{2}^{2} \alpha_{3}^{2}(1+\alpha_{7})/\alpha_{6}$, and  
$d_{44}=(\beta_{2}-\beta_{1})^{2}\alpha_{1}^{2} \alpha_{4}^{2}(1+\alpha_{8})/\alpha_{5}$.
Here in order for $\gamma$ to be positive definite, we require $\beta_{1}\ne \beta_{2}$, $\beta_{1}\beta_{2}\ne -1$, $\alpha_{1}\ne 0$, $\alpha_{2}\ne 0$, $\alpha_{3}\ne 0$, $\alpha_{4}\ne 0$, $\alpha_{5}> 0$, $\alpha_{6}> 0$, $\alpha_{7}> 0$, and  $\alpha_{8}> 0$. Then a direct calculation yields
\begin{align}
\gamma_{11}&=\alpha_{5}+(1+\beta_{1}^{2})\alpha_{1}^{2},  \label{r11}\\
\gamma_{22}&
=\frac{1+\alpha_{8}}{\alpha_{5}\alpha_{8}},  \label{r22}\\
\gamma_{33}&=\alpha_{6}+(1+\beta_{1}^{2})\alpha_{2}^{2},  \label{r33}\\
\gamma_{44}&
=\frac{1+\alpha_{7}}{\alpha_{6}\alpha_{7}},  \label{r44}\\
\gamma_{55}&=(\beta_{1}-\beta_{2})^{2}\alpha_{2}^{2} \alpha_{3}^{2}(1+\alpha_{7})/\alpha_{6}+(1+\beta_{2}^{2})\alpha_{3}^{2},  \label{r55}\\
\gamma_{66}&=\frac{\alpha_{6}}{\alpha_{7}(\beta_{1}-\beta_{2})^{2}\alpha_{2}^{2} \alpha_{3}^{2}},  \label{r66}\\
\gamma_{77}&=(\beta_{2}-\beta_{1})^{2}\alpha_{1}^{2} \alpha_{4}^{2}(1+\alpha_{8})/\alpha_{5}+(1+\beta_{2}^{2})\alpha_{4}^{2},  \label{r77}\\
\gamma_{88}&=\frac{\alpha_{5}}{\alpha_{8}(\beta_{2}-\beta_{1})^{2}\alpha_{1}^{2} \alpha_{4}^{2}},  \label{r88}\\
\gamma_{15}&=(1+\beta_{1}\beta_{2})\alpha_{1} \alpha_{3},  \label{r15}\\
\gamma_{28}&
=\frac{1}{(\beta_{1}-\beta_{2})\alpha_{1}\alpha_{4}\alpha_{8}},  \label{r28}\\
\gamma_{37}&=(1+\beta_{1}\beta_{2})\alpha_{2} \alpha_{4}, \label{r37}\\
\gamma_{46}&
=\frac{1}{(\beta_{2}-\beta_{1})\alpha_{2}\alpha_{3}\alpha_{7}}.  \label{r46}
\end{align}
 
\begin{thm}\label{thm}
For any real $\beta_{1}\ne \beta_{2}$,  $\beta_{1}\beta_{2}\ne -1$, $\alpha_{1}\ne 0$, $\alpha_{2}\ne 0$, $\alpha_{3}\ne 0$, $\alpha_{4}\ne 0$, $\alpha_{5}> 0$, $\alpha_{6}> 0$, $\alpha_{7}> 0$, and  $\alpha_{8}> 0$, a matrix of the form~\eqref{r} with entries determined by Eqs.~\eqref{r11} -~\eqref{r46} always corresponds to a $2\times 2$ bound entangled Gaussian state.
\end{thm}

\emph{Examples of Bound Entangled States.}
We construct four examples of  $2\times 2$ bound entangled Gaussian states according to the parametrization described in Theorem~\ref{thm}. The first example is already shown in Ref.~\cite{WW01:prl}. It is demonstrated that this bound entangled state can also be obtained using the parametrization above. 

\emph{Example 1.} Choose $\beta_{1}=1$, $\beta_{2}=2$, $\alpha_{1}=-\alpha_{2}=\alpha_{3}=\alpha_{4}=\frac{\sqrt{3}}{3}$, $\alpha_{5}=\alpha_{6}=\frac{4}{3}$, and $\alpha_{7}=\alpha_{8}=3$. The resulting covariance matrix calculated from Eqs. ~\eqref{r11} -~\eqref{r46} is 
\begin{align}
\gamma=\begin{pmatrix}
   2 &0 &0 &0 &1 &0 &0 &0\\
   0 &1 &0 &0 &0 &0 &0 &-1\\
   0 &0 &2 &0 &0 &0 &-1 &0\\
   0 &0 &0 &1 &0 &-1 &0 &0\\
   1 &0 &0 &0 &2 &0 &0 &0\\
   0 &0 &0 &-1 &0 &4 &0 &0\\
   0 &0 &-1 &0 &0 &0 &2 &0\\
   0 &-1 &0 &0 &0 &0 &0 &4
\end{pmatrix}. \label{WWpaper}
\end{align}

\emph{Example 2.} Choose $\beta_{1}=1$, $\beta_{2}=3$, $\alpha_{1}=\alpha_{2}=\alpha_{3}=\alpha_{4}=\frac{\sqrt{2}}{2}$, $\alpha_{5}=\alpha_{6}=\alpha_{7}=\alpha_{8}=1$. The resulting covariance matrix calculated from Eqs. ~\eqref{r11} -~\eqref{r46} is 
\begin{align}
\gamma=\begin{pmatrix}
   2 &0 &0 &0 &2 &0 &0 &0\\
   0 &2 &0 &0 &0 &0 &0 &-1\\
   0 &0 &2 &0 &0 &0 &2 &0\\
   0 &0 &0 &2 &0 &1 &0 &0\\
   2 &0 &0 &0 &7 &0 &0 &0\\
   0 &0 &0 &1 &0 &1 &0 &0\\
   0 &0 &2 &0 &0 &0 &7 &0\\
   0 &-1 &0 &0 &0 &0 &0 &1
\end{pmatrix}. \label{exam2}
\end{align} 

\emph{Example 3.} Choose $\beta_{1}=1/3$, $\beta_{2}=1/2$, $\alpha_{1}=\alpha_{2}=3/2$, $\alpha_{3}=\alpha_{4}=4$, $\alpha_{5}=\alpha_{6}=\alpha_{7}=\alpha_{8}=1/2$. The resulting covariance matrix calculated from Eqs. ~\eqref{r11} -~\eqref{r46} is 
\begin{align}
\gamma=\begin{pmatrix}
   3 &0 &0 &0 &7 &0 &0 &0\\
   0 &6 &0 &0 &0 &0 &0 &-2\\
   0 &0 &3 &0 &0 &0 &7 &0\\
   0 &0 &0 &6 &0 &2 &0 &0\\
   7 &0 &0 &0 &23 &0 &0 &0\\
   0 &0 &0 &2 &0 &1 &0 &0\\
   0 &0 &7 &0 &0 &0 &23 &0\\
   0 &-2 &0 &0 &0 &0 &0 &1
\end{pmatrix}. \label{exam3}
\end{align} 

\emph{Example 4.} Choose $\beta_{1}=-\sqrt{2}$, $\beta_{2}=2\sqrt{2}$, $\alpha_{1}=1/2$, $\alpha_{2}=-\sqrt{2}/2$, $\alpha_{3}=1/3$, $\alpha_{4}=-\sqrt{2}/2$, $\alpha_{5}=1$, $\alpha_{6}=3$, $\alpha_{7}=2$, $\alpha_{8}=2/9$. The resulting covariance matrix calculated from Eqs. ~\eqref{r11} -~\eqref{r46} is 
\begin{align}
\gamma=\begin{pmatrix}
   \frac{7}{4} &0 &0 &0 &-\frac{1}{2} &0 &0 &0\\
   0 &\frac{11}{2} &0 &0 &0 &0 &0 &3\\
   0 &0 &\frac{9}{2} &0 &0 &0 &-\frac{3}{2} &0\\
   0 &0 &0 &\frac{1}{2} &0 &-\frac{1}{2} &0 &0\\
   -\frac{1}{2} &0 &0 &0 &2 &0 &0 &0\\
   0 &0 &0 &-\frac{1}{2} &0 &\frac{3}{2} &0 &0\\
   0 &0 &-\frac{3}{2} &0 &0 &0 &\frac{29}{4} &0\\
   0 &3 &0 &0 &0 &0 &0 &2
\end{pmatrix}. \label{exam4}
\end{align} 
The covariance matrices~\eqref{WWpaper},~\eqref{exam2},~\eqref{exam3} and \eqref{exam4} are all $2\times 2$ bound entangled states. In particular, the covariance matrix \eqref{exam4} \emph{cannot} be constructed using the method developed in Ref.~\cite{WW01:prl}. This is because the matrix \eqref{exam4} does not commutate with the skew symmetric matrix $R$ with $R_{13}=R_{24}=R_{75}=R_{86}=1$, and zero remaining entries as defined in Ref.~\cite{WW01:prl}. 

\emph{Preparation of the bound entangled state \eqref{WWpaper}.} We  propose an experimental protocol for generating the bound entangled state with covariance matrix \eqref{WWpaper}.  According to Williamson's theorem~\cite{W36:ajm}, the covariance matrix~\eqref{WWpaper} can be diagonalized through a symplectic transformation. It is found that $\gamma=SDS^{T}$, 
where $D=\diag(1,1,1,1,3,3,3,3)$ contains the symplectic eigenvalues of $\gamma$ and the symplectic matrix $S$ satisfies $S\sigma S^{T}=\sigma$ and is given by 
\begin{align}
S&=\begin{pmatrix}
0 &s_{12} &0 &s_{14} &0 &s_{16} &0 &s_{18}\\
s_{21} &0 &s_{23} &0 &s_{25} &0 &s_{27} &0\\
0 &s_{14} &0 &-s_{12} &0 &s_{18} &0 &-s_{16}\\
s_{23} &0 &-s_{21} &0 &s_{27} &0 &-s_{25} &0\\
0 &s_{52} &0 &s_{54} &0 &s_{56} &0 &s_{58}\\
s_{83} &0 &-s_{81} &0 &s_{87} &0 &-s_{85} &0\\
0 &-s_{54} &0 &s_{52} &0 &-s_{58} &0 &s_{56}\\
s_{81} &0 &s_{83} &0 &s_{85} &0 &s_{87} &0
\end{pmatrix}, \label{exam_S}
\end{align}
with
\begin{align*}
s_{12}&=\frac{(-\sqrt{13}-3)\sqrt{5+\sqrt{13}}+(3-\sqrt{13})\sqrt{5-\sqrt{13}}}{8\sqrt{13}},\\
s_{14}&=\frac{(\sqrt{39}+4\sqrt{3})\sqrt{5+\sqrt{13}}+(\sqrt{39}-4\sqrt{3})\sqrt{5-\sqrt{13}}}{12\sqrt{13}},\\
s_{16}&=\frac{(\sqrt{39}+3\sqrt{3})\sqrt{5+\sqrt{13}}+(\sqrt{39}-3\sqrt{3})\sqrt{5-\sqrt{13}}}{8\sqrt{7}\sqrt{13}},\\
s_{18}&=\frac{(4-\sqrt{13})\sqrt{5+\sqrt{13}}-(4+\sqrt{13})\sqrt{5-\sqrt{13}}}{4\sqrt{7}\sqrt{13}}, \\
s_{21}&=\frac{(\sqrt{39}-3\sqrt{3})\sqrt{5+\sqrt{13}}+(\sqrt{39}+3\sqrt{3})\sqrt{5-\sqrt{13}}}{8\sqrt{13}} ,\\
s_{23}&=\frac{(4-\sqrt{13})\sqrt{5+\sqrt{13}}-(4+\sqrt{13})\sqrt{5-\sqrt{13}}}{4\sqrt{13}} ,\\
s_{25}&=\frac{(3-\sqrt{13})\sqrt{5+\sqrt{13}}-(3+\sqrt{13})\sqrt{5-\sqrt{13}}}{8\sqrt{7}\sqrt{13}},\\
s_{27}&=\frac{(\sqrt{13}+4)\sqrt{5+\sqrt{13}}+(\sqrt{13}-4)\sqrt{5-\sqrt{13}}}{4\sqrt{3}\sqrt{7}\sqrt{13}},\\
s_{52}&=\frac{(\sqrt{39}+\sqrt{3})\sqrt{5+\sqrt{13}}+(\sqrt{39}-\sqrt{3})\sqrt{5-\sqrt{13}}}{24\sqrt{13}},\\
s_{54}&=\frac{ \sqrt{5+\sqrt{13}}-\sqrt{5-\sqrt{13}}}{4\sqrt{13}},\\
s_{56}&=\frac{ (7\sqrt{13}-25)\sqrt{5+\sqrt{13}}+(7\sqrt{13}+25)\sqrt{5-\sqrt{13}}}{8\sqrt{7}\sqrt{13}},\\
s_{58}&=\frac{ -\sqrt{3}\sqrt{5+\sqrt{13}}+\sqrt{3}\sqrt{5-\sqrt{13}}}{4\sqrt{7}\sqrt{13}} ,\\
s_{81}&=\frac{-\sqrt{3}\sqrt{5+\sqrt{13}}+\sqrt{3}\sqrt{5-\sqrt{13}}}{4\sqrt{13} },\\
s_{83}&=\frac{(-1+\sqrt{13})\sqrt{5+\sqrt{13}}+(1+\sqrt{13})\sqrt{5-\sqrt{13}}}{-8\sqrt{13} },\\
s_{85}&=\frac{\sqrt{5+\sqrt{13}}-\sqrt{5-\sqrt{13}}}{4\sqrt{7}\sqrt{13} },\\
s_{87}&=\frac{(-25-7\sqrt{13})\sqrt{5+\sqrt{13}}+(25-7\sqrt{13})\sqrt{5-\sqrt{13}}}{8\sqrt{3}\sqrt{7}\sqrt{13}}.
\end{align*}
Here the symplectic eigenvalues of $\gamma$ can also be computed from the standard eigenspectrum of the
matrix $i\sigma \gamma$. Using the Euler decomposition~\cite{ADMS95:Pramana,B05:pra}, the symplectic matrix $S$ in Eq.~\eqref{exam_S} can be further decomposed as  
\begin{align}
S=K[\bigoplus\limits_{k=1}^{4} S(r_{k})]L, \label{Eular}
\end{align}
where $K$ and $L$ are symplectic and orthogonal matrices and correspond to passive canonical unitaries (i.e., the ones that preserve the average photon number of the input state), while $S(r_{1})$, $\cdots$, $S(r_{4})$ is a set of one-mode squeezing matrices. We find  \begin{align}
&S(r_{1})=S(r_{2})=S(r_{3})=S(r_{4})=\begin{pmatrix}
\frac{\sqrt{17}+1}{4} &0\\
0 &\frac{\sqrt{17}-1}{4}
\end{pmatrix}. \label{squeezing}
\end{align}
The values of the matrices $K$ and $L$ can be found in~\cite{Supplement}. 
Therefore, the bound entangled Gaussian state $\gamma$ in Eq.~\eqref{WWpaper} can be decomposed as 
\begin{align}
\gamma=K[\bigoplus\limits_{k=1}^{4} S(r_{k})]L D L^{T} [\bigoplus\limits_{k=1}^{4} S(r_{k})] K^{T}. \label{decomposition}
\end{align}
Thus the bound entangled Gaussian state \eqref{WWpaper} can be prepared beginning with an initial product state corresponding to the diagonal matrix $D$, and then applying a  multiport interferometer $L$, a parallel set of one-mode squeezers  $S(r_{k})$ and finally a multiport interferometer  $K$. 
The transformations described by the multiport interferometers $L$ and $K$  can  be both implemented using a network of beam splitters and phase shifters~\cite{RZBB94:prl,CHMKW16:opt}; see \cite{Supplement} for details.  

\begin{figure}[htbp]
\begin{center}
\includegraphics[height=3.15cm]{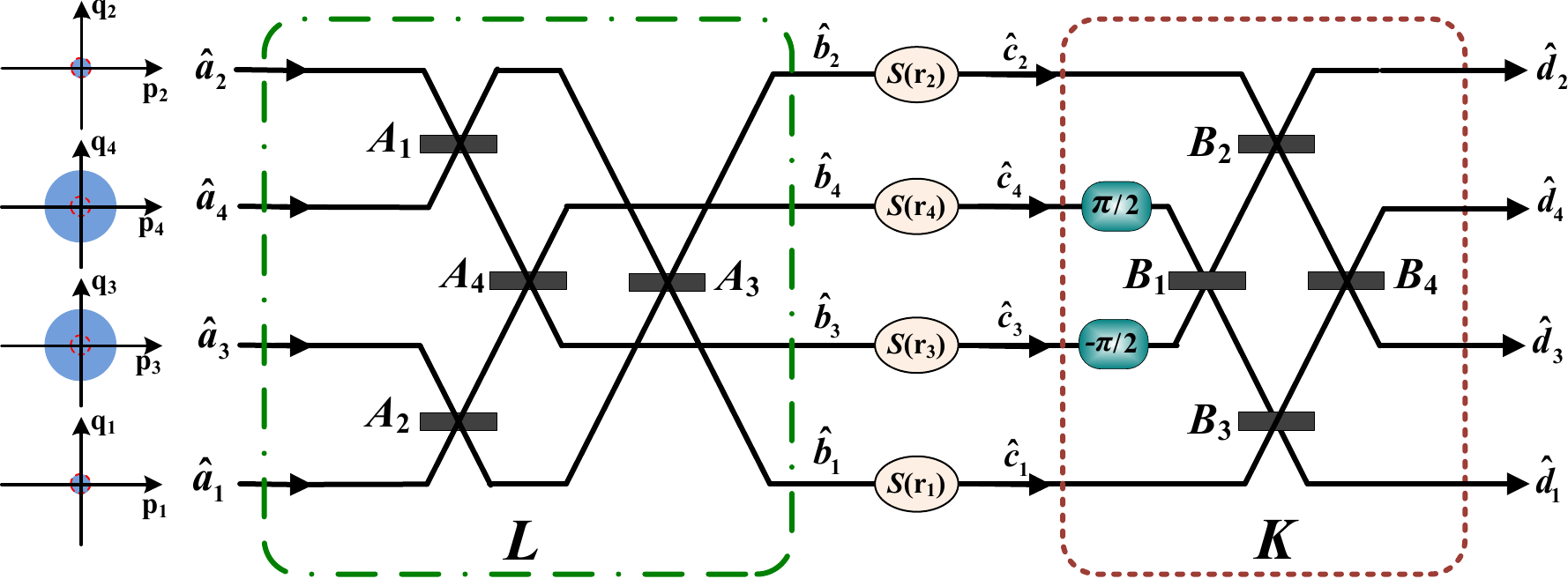}
\caption{A diagram for preparation of the bound entangled Gaussian state \eqref{WWpaper}. The initial states $\hat{a}_{1}$ and $\hat{a}_{2}$ are  in a vacuum state while the other two optical modes $\hat{a}_{3}$ and $\hat{a}_{4}$ are in a thermal state with covariance matrix $3I$ (i.e., the average photon number of the thermal field is $\bar{n}=1$). The unitary multiport interferometers  $L$ and $K$ are realized via using a network of beam splitters and phase shifters. The transmittances and reflectances of the beam splitters $A_{1},\cdots, A_{4}$ and $B_{1},\cdots,B_{4}$ are determined by the corresponding unitary transformations; see \cite{Supplement} for details.  
The box
labeled $-\pi/2$ ($\pi/2$) represents the
relative phase shift $\hat{a}\to i \hat{a}$ ($-i\hat{a}$). 
$S(r_{1})$, $\cdots$, $S(r_{4})$ are a set of one-mode squeezers as described by Eq.~\eqref{squeezing}. The output Gaussian state ($\hat{d}_{1},\cdots,\hat{d}_{4}$) has the covariance matrix \eqref{WWpaper}, and is a bound entangled state with respect to the bipartite splitting such that Alice possesses modes $\{\hat{d}_{1},\hat{d}_{2}\}$ and Bob possesses modes $\{\hat{d}_{3},\hat{d}_{4}\}$.}
\label{all_realize}
\end{center}
\end{figure}

Combining the  above analysis, the experimental protocol for preparing the  bound entangled Gaussian state \eqref{WWpaper} is depicted in Fig.~\ref{all_realize}. The input fields $\hat{a}_{1}$ and $\hat{a}_{2}$ are in the vacuum state while other two input fields $\hat{a}_{3}$ and $\hat{a}_{4}$  are in the thermal state with covariance matrix $3I$ (i.e., the average photon number is $\bar{n}=1$). By applying a multiport interferometer $L$, a parallel set of one-mode squeezers  $S(r_{k})$ and finally a multiport interferometer  $K$, the Gaussian state obtained at the output ($\hat{d}_{1},\cdots,\hat{d}_{4}$) has the covariance matrix \eqref{WWpaper}, and is a bound entangled state with respect to the bipartite splitting such that Alice possesses modes $\{\hat{d}_{1},\hat{d}_{2}\}$ and Bob possesses modes $\{\hat{d}_{3},\hat{d}_{4}\}$.

\begin{figure}[htbp]
\begin{center}
\includegraphics[height=4.8cm]{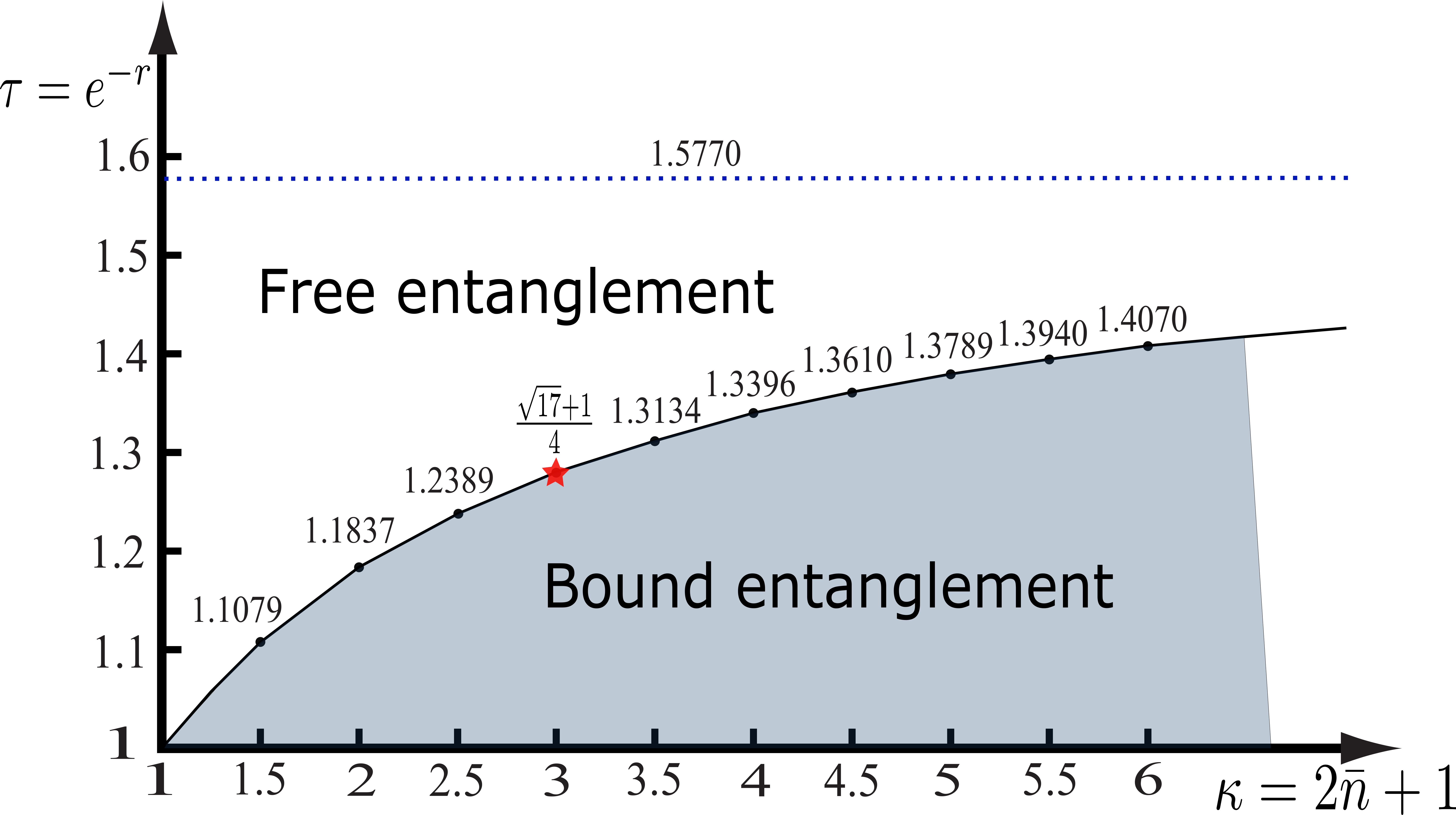}
\caption{The  bound entanglement region obtained from the system described in Fig.~\ref{all_realize} by  varying the thermal inputs and the squeezing components. The shaded region corresponds to the bound entangled states with respect to the bipartition $\{\{\hat{d}_{1},\hat{d}_{2}\},\{\hat{d}_{3},\hat{d}_{4}\}\}$.  The point marked with a red star corresponds to the bound entangled state~\eqref{WWpaper} which is generated  when $\kappa=3$ and $\tau=(\sqrt{17}+1)/4$. It lies on the boundary between bound and free entangled states. As the average photon number $\bar{n}$ increases, the boundary curve between the bound and free entanglement regions approaches a horizontal asymptote $\tau=1.5770$ (marked by a dotted blue line). Thus when the squeezing parameter $r$ satisfies $e^{-r}>1.5770$, we will always obtain free entanglement at the output no matter how large the value of $\bar{n}$.
}
\label{bound1133_12to34}
\end{center}
\end{figure}

\emph{Bound Entanglement Region.} Consider the optical system
depicted in Fig.~\ref{all_realize}. We fix the passive unitaries $L$ and $K$ (i.e., the corresponding beam splitters and phase shifters implementing $L$ and $K$ remain unchanged). Also, we fix the optical inputs $\hat{a}_{1}$ and $\hat{a}_{2}$ which remain in the vacuum. Suppose the other  optical inputs $\hat{a}_{3}$ and $\hat{a}_{4}$ are in the same thermal state with covariance matrix $(2\bar{n}+1)I$ where $\bar{n}$ is the average photon number. Suppose the squeezers $S(r_{1})$, $\cdots$, $S(r_{4})$ between $L$ and $K$ realize the same symplectic transformation; that is,  $S(r_{1})=S(r_{2})=S(r_{3})=S(r_{4})=\begin{pmatrix}
e^{-r} &0\\
0 &e^{r}
\end{pmatrix}$, where $r\in\mathbb{R}$ is the squeezing parameter. As discussed before, if $\kappa:=2\bar{n}+1=3$ and $\tau:=e^{-r}=(\sqrt{17}+1)/4$, the state produced at the output has the covariance matrix \eqref{WWpaper}, and is bound entangled  with respect to the bipartition $\{\{\hat{d}_{1},\hat{d}_{2}\},\{\hat{d}_{3},\hat{d}_{4}\}\}$. Now we vary the thermal inputs $\bar{n}$ and the squeezing parameter $r$ such that we can obtain different Gaussian states at the output. The entanglement properties of these output states can be determined from their covariance matrices via solving a semidefinite programming problem. The results are shown in Fig.~\ref{bound1133_12to34}. As can be seen in Fig.~\ref{bound1133_12to34}, without the presence of squeezing,  no entanglement can be generated. If we add a small amount of squeezing, the output state should be bound entangled. However, if we continue to increase the amount of squeezing, the output state should eventually enter a region of free entanglement.

In conclusion, we have parametrized a class of $2\times 2$ bound entangled Gaussian states. For a particular bound entangled state, we present an experimental protocol for generating it in quantum optics.  It is interesting to extend this result to continuous-variable  multipartite bound entangled states, which may serve as a useful  resource for multiparty quantum communication  such as remote information concentration~\cite{MV01:prl}, secure quantum key distribution \cite{HHHO05:prl,AH09:pra}, and superactivation~\cite{BR05:pra}. We believe the results we present here  may contribute to a deeper understanding of entanglement in the continuous-variable domain.


\clearpage

\onecolumngrid
\begin{center}
\vspace*{\baselineskip}
{\textbf{\large Supplemental Material}}\\
\end{center}

\renewcommand{\theequation}{S\arabic{equation}}
\setcounter{equation}{0}
\setcounter{figure}{0}
\setcounter{table}{0}
\setcounter{section}{0}
\setcounter{page}{1}
\makeatletter

\section{The matrices $K$ and $L$}
The unitary matrices $K$ and $L$ appearing in Eq.~\eqref{Eular} are given by 
\begin{align*}
&K=\frac{1}{\sqrt{17-3\sqrt{17}}}\begin{pmatrix}      
2          &0            &0         &0          &0           &-\frac{\sqrt{17}-3}{2}     &0         &-\frac{\sqrt{17}-3}{2}\\  
0          &2            &0         &0           &\frac{\sqrt{17}-3}{2}           & 0         &\frac{\sqrt{17}-3}{2}     &     0\\
0          &0           &2           &0             &0           &-\frac{\sqrt{17}-3}{2}      &0    & \frac{\sqrt{17}-3}{2}\\
0          &0           &0            &2                 &\frac{\sqrt{17}-3}{2}       &0        &-\frac{\sqrt{17}-3}{2}     &      0 \\
\frac{\sqrt{17}-3}{2}       &0           &\frac{\sqrt{17}-3}{2}    &0       &0                  &2         &0          &0 \\
0        &\frac{\sqrt{17}-3}{2}     &0           &\frac{\sqrt{17}-3}{2}                 &-2                  &0            &0     &0 \\
\frac{\sqrt{17}-3}{2}      &0        &-\frac{\sqrt{17}-3}{2}  &0       &0                  &0           &0           &  2 \\
0        &\frac{\sqrt{17}-3}{2}     &0            &-\frac{\sqrt{17}-3}{2}           &0             & 0        &-2            &0 \end{pmatrix},\\
&L=\frac{1}{\sqrt{17-3\sqrt{17}}}\begin{pmatrix}
0 &l_{12} &0 &l_{14} &0 &l_{16} &0 &l_{18}\\
-l_{12} &0 &-l_{14} &0 &-l_{16} &0 &-l_{18} &0\\
0  &l_{14} &0 &-l_{12} &0 &l_{18} &0 &-l_{16}\\
-l_{14} &0 &l_{12} &0 &-l_{18} &0 &l_{16} &0\\ 
l_{51} &0 &l_{53} &0 &l_{55} &0 &l_{57} &0\\ 
0 &l_{51} &0 &l_{53} &0 &l_{55} &0 &l_{57}\\
-l_{53} &0 &l_{51} &0 &-l_{57} &0 &l_{55} &0\\ 
0 &-l_{53} &0 &l_{51} &0 &-l_{57} &0 &l_{55}
\end{pmatrix},
\end{align*}
\begin{align*}
l_{12}=&
\frac{ \Big( -(21+3\sqrt{17})  +  3(1-\sqrt{17})\sqrt{13}  +  (5-\sqrt{17})\sqrt{39}  +  (5-\sqrt{17})\sqrt{3} \Big) \sqrt{5+\sqrt{13}}}{48\sqrt{13}} \\
&+
\frac{ \Big(   (21+3\sqrt{17})+3(1-\sqrt{17})\sqrt{13}+(5-\sqrt{17})\sqrt{39}-(5-\sqrt{17})\sqrt{3}  \Big)\sqrt{5-\sqrt{13}}}{48\sqrt{13}},\\
l_{14}=& \frac{\Big((30-6\sqrt{17})+(3+\sqrt{17})\sqrt{39}+(7\sqrt{17}-3)\sqrt{3}\Big)\sqrt{5+\sqrt{13}}}{48\sqrt{13}} \\
&+\frac{ \Big(-(30-6\sqrt{17})+(3+\sqrt{17})\sqrt{39}-(7\sqrt{17}-3)\sqrt{3}\Big)\sqrt{5-\sqrt{13}}}{48\sqrt{13}},\\
l_{16}=& \frac{\Big((35-7\sqrt{17})\sqrt{13}-(125-25\sqrt{17})+(\sqrt{17}-1)\sqrt{39}+(7+\sqrt{17})\sqrt{3}\Big)\sqrt{5+\sqrt{13}}}{16\sqrt{7}\sqrt{13}} \\
&+ \frac{\Big((35-7\sqrt{17})\sqrt{13}+(125-25\sqrt{17})+(\sqrt{17}-1)\sqrt{39}-(7+\sqrt{17})\sqrt{3}\Big)\sqrt{5-\sqrt{13}}}{16\sqrt{7}\sqrt{13}},    \\
l_{18}= & \frac{\Big((37-9\sqrt{17})\sqrt{13}+(33\sqrt{17}-133)+(2\sqrt{17}-10)\sqrt{3}\Big)\sqrt{5+\sqrt{13}}}{16\sqrt{7}\sqrt{13}} \\
&+ \frac{\Big((37-9\sqrt{17})\sqrt{13}+(133-33\sqrt{17})+(10-2\sqrt{17})\sqrt{3}\Big)\sqrt{5-\sqrt{13}}}{16\sqrt{7}\sqrt{13} },  \\
l_{51}= &\frac{ \Big((4\sqrt{17}-12)\sqrt{39}+(10\sqrt{17}-54)\sqrt{3}+(63-9\sqrt{17})+(21-3\sqrt{17})\sqrt{13}\Big)\sqrt{5+\sqrt{13}}}{96\sqrt{13}} \\
&+  \frac{\Big((4\sqrt{17}-12)\sqrt{39}-(10\sqrt{17}-54)\sqrt{3}-(63-9\sqrt{17})+(21-3\sqrt{17})\sqrt{13}\Big)\sqrt{5-\sqrt{13}}}{96\sqrt{13}},   \\
l_{53}= &\frac{ \Big((21\sqrt{17}-51)+(3\sqrt{17}-21)\sqrt{13}+(2\sqrt{17}-14)\sqrt{39}+(8\sqrt{17}-56)\sqrt{3}\Big)\sqrt{5+\sqrt{13}}} {96\sqrt{13}}\\
&+ \frac{\Big((-21\sqrt{17}+51)+(3\sqrt{17}-21)\sqrt{13}+(2\sqrt{17}-14)\sqrt{39}-(8\sqrt{17}-56)\sqrt{3}\Big)\sqrt{5-\sqrt{13}}}{96\sqrt{13}}  , \\
l_{55}=& \frac{\Big(-(106+42\sqrt{17})+(12\sqrt{17}+28)\sqrt{13}+(\sqrt{17}-7)\sqrt{39}+(3\sqrt{17}-21)\sqrt{3}\Big)\sqrt{5+\sqrt{13}}}{32\sqrt{7}\sqrt{13}}\\
& + \frac{\Big((106+42\sqrt{17})+(28+12\sqrt{17})\sqrt{13}+(21-3\sqrt{17})\sqrt{3}+(\sqrt{17}-7)\sqrt{39}\Big)\sqrt{5-\sqrt{13}}}{32\sqrt{7}\sqrt{13}}, \\
l_{57}=& \frac{\Big((-56+8\sqrt{17})+(14-2\sqrt{17})\sqrt{13}+(7-\sqrt{17})\sqrt{39}+(17-7\sqrt{17})\sqrt{3}\Big)\sqrt{5+\sqrt{13}}}{32\sqrt{7}\sqrt{13}}\\
& + \frac{\Big((56-8\sqrt{17})+(14-2\sqrt{17})\sqrt{13}+(7-\sqrt{17})\sqrt{39}-(17-7\sqrt{17})\sqrt{3}\Big)\sqrt{5-\sqrt{13}}}{32\sqrt{7}\sqrt{13}}.
\end{align*}

\section{Realization of the transformations $L$ and $K$}
The input-output relations described by $L$ and $K$ can be, respectively, written as
\begin{align*}
\begin{pmatrix}
\hat{b}_{1}\\
\hat{b}_{2}\\
\hat{b}_{3}\\
\hat{b}_{4}
\end{pmatrix}&=\frac{1}{\sqrt{17-3\sqrt{17}}}\begin{pmatrix}
-il_{12} &-il_{14} &-il_{16} &-il_{18}\\
-il_{14} &il_{12} &-il_{18} &il_{16}\\
l_{51} &l_{53} &l_{55} &l_{57}\\
-l_{53} &l_{51} &-l_{57} &l_{55}
\end{pmatrix}\begin{pmatrix}
\hat{a}_{1}\\
\hat{a}_{2}\\
\hat{a}_{3}\\
\hat{a}_{4}
\end{pmatrix},\\
\begin{pmatrix}
\hat{d}_{1}\\
\hat{d}_{2}\\
\hat{d}_{3}\\
\hat{d}_{4}
\end{pmatrix}&=\frac{1}{\sqrt{17-3\sqrt{17}}}\begin{pmatrix}
2 &0 &\frac{\sqrt{17}-3}{2}i &\frac{\sqrt{17}-3}{2}i\\
0 &2 &\frac{\sqrt{17}-3}{2}i &-\frac{\sqrt{17}-3}{2}i\\
\frac{\sqrt{17}-3}{2} &\frac{\sqrt{17}-3}{2} &-2i &0\\
\frac{\sqrt{17}-3}{2} &-\frac{\sqrt{17}-3}{2} &0 &-2i
\end{pmatrix}\begin{pmatrix}
\hat{c}_{1}\\
\hat{c}_{2}\\
\hat{c}_{3}\\
\hat{c}_{4}
\end{pmatrix}.
\end{align*}

\notag \emph{Realization of the multiport interferometer $L$:} By direct calculation, we find 
\begin{align*}
\frac{1}{\sqrt{17-3\sqrt{17}}}\begin{pmatrix}
-il_{12} &-il_{14} &-il_{16} &-il_{18}\\
-il_{14} &il_{12} &-il_{18} &il_{16}\\
l_{51} &l_{53} &l_{55} &l_{57}\\
-l_{53} &l_{51} &-l_{57} &l_{55}
\end{pmatrix}=A_{4}A_{3}A_{2}A_{1}A_{0}\begin{pmatrix}
1 &0 &0 &0\\
0 &-1 &0 &0\\
0 &0 &1 &0\\
0 &0 &0 &1
\end{pmatrix},
\end{align*}
where 
\begin{align}
A_{0}&=\begin{pmatrix}
1 &0 &0 &0\\
0 &1 &0 &0\\
0 &0 &\frac{l_{51} l_{55}+l_{57} l_{53}}{\sqrt{(l_{51}^2+l_{53}^2) (l_{55}^2+l_{57}^2)}}  &\frac{l_{57} l_{51}-l_{53} l_{55}}{\sqrt{(l_{51}^2+l_{53}^2) (l_{55}^2+l_{57}^2)}}\\
        0 &0 &\frac{l_{53} l_{55}-l_{57} l_{51}}{\sqrt{(l_{51}^2+l_{53}^2) (l_{55}^2+l_{57}^2)}}  &\frac{l_{51} l_{55}+l_{53} l_{57}}{\sqrt{(l_{51}^2+l_{53}^2) (l_{55}^2+l_{57}^2)}  }
\end{pmatrix}, \label{beam11}\\
A_{1}&=\frac{1}{\sqrt{17-3\sqrt{17}}}\begin{pmatrix}
1  &0 &0 &0\\
0 &-i\sqrt{l_{12}^2+l_{14}^2} &0  &-i\sqrt{l_{51}^2+l_{53}^2}\\
0  &0 &1    &0 \\
0  &-\sqrt{l_{51}^2+l_{53}^2} &0  &\sqrt{l_{12}^2+l_{14}^2 }
\end{pmatrix}, \label{beam12} \\
A_{2}&=\frac{1}{\sqrt{17-3\sqrt{17}}}\begin{pmatrix}
-i\sqrt{l_{12}^2+l_{14}^2} &0  &i\sqrt{l_{51}^2+l_{53}^2} &0\\
 0   &1    &0   &0\\
 \sqrt{l_{51}^2+l_{53}^2} &0  &\sqrt{l_{12}^2+l_{14}^2} &0\\
 0 &0 &0 &1 
\end{pmatrix}, \label{beam13} \\
A_{3}&=\begin{pmatrix}
\frac{l_{12}}{\sqrt{l_{12}^2+l_{14}^2}}  &-\frac{l_{14}}{\sqrt{l_{12}^2+l_{14}^2}} &0 &0\\
\frac{l_{14}}{\sqrt{l_{12}^2+l_{14}^2}} &\frac{l_{12}}{\sqrt{l_{12}^2+l_{14}^2}} &0 &0\\
0 &0 &1 &0\\
        0 &0 &0 &1  
\end{pmatrix},\;\;
A_{4}=\begin{pmatrix}
1 &0 &0 &0\\
0 &1 &0 &0\\
0 &0 &\frac{l_{51}}{\sqrt{l_{51}^2+l_{53}^2}}  &\frac{l_{53}}{\sqrt{l_{51}^2+l_{53}^2}}\\
0 &0 &-\frac{l_{53}}{\sqrt{l_{51}^2+l_{53}^2}}  &\frac{l_{51}}{\sqrt{l_{51}^2+l_{53}^2}} 
\end{pmatrix}. \label{beam14}
\end{align}
are Bogoliubov transformations and can be realized using beam splitters. The matrix $\begin{pmatrix}
1 &0 &0 &0\\
0 &-1 &0 &0\\
0 &0 &1 &0\\
0 &0 &0 &1
\end{pmatrix}$  implements a $\pi$ phase shift onto the optical input $\hat{a}_{2}$. However, this $\pi$ phase shift is not relevant in our case since the optical input field $\hat{a}_{2}$ is in the vacuum. Also, the beam splitter $A_{0}$ is not relevant in our case since it acts on two identical thermal inputs $\hat{a}_{3}$ and $\hat{a}_{4}$. It makes no difference whether these two components (the $\pi$ phase shifter and the beam splitter $A_{0}$) are added to the quantum system or not. Therefore, they are removed and do not appear in Fig.~\ref{all_realize}.  \\

\notag \emph{Realization of the multiport interferometer $K$:} By direct calculation, we find  
\begin{align*}
\frac{1}{\sqrt{17-3\sqrt{17}}}\begin{pmatrix}
2 &0 &\frac{\sqrt{17}-3}{2}i &\frac{\sqrt{17}-3}{2}i\\
0 &2 &\frac{\sqrt{17}-3}{2}i &-\frac{\sqrt{17}-3}{2}i\\
\frac{\sqrt{17}-3}{2} &\frac{\sqrt{17}-3}{2} &-2i &0\\
\frac{\sqrt{17}-3}{2} &-\frac{\sqrt{17}-3}{2} &0 &-2i
\end{pmatrix}=B_{4}B_{3}B_{2}B_{1}\begin{pmatrix}
1 &0 &0 &0\\
0 &1 &0 &0\\
0 &0 &i &0\\
0 &0 &0 &-i
\end{pmatrix},
\end{align*} where 
\begin{align}
B_{1}&=\begin{pmatrix}
1 &0 &0 &0\\
0 &1 &0 &0\\
0 &0 &-\frac{\sqrt{2}}{2} &\frac{\sqrt{2}}{2}\\
0 &0 &-\frac{\sqrt{2}}{2} &-\frac{\sqrt{2}}{2}
\end{pmatrix},
\quad 
B_{2}=\begin{pmatrix}
1 &0 &0 &0\\
0 &\frac{2}{\sqrt{17-3\sqrt{17}}} &0 &\frac{-(\sqrt{17}-3)}{\sqrt{2}\sqrt{17-3\sqrt{17}}} \\
0 &0 &1 &0 \\
0 &\frac{(\sqrt{17}-3)}{\sqrt{2}\sqrt{17-3\sqrt{17}}} &0 &\frac{2}{\sqrt{17-3\sqrt{17}}} \\
\end{pmatrix}, \label{beam21}\\ 
B_{3}&=\begin{pmatrix}
\frac{2}{\sqrt{17-3\sqrt{17}}} &0 &\frac{-(\sqrt{17}-3)}{\sqrt{2}\sqrt{17-3\sqrt{17}}} &0\\
0 &1 &0 &0\\
\frac{(\sqrt{17}-3)}{\sqrt{2}\sqrt{17-3\sqrt{17}}} &0 &\frac{2}{\sqrt{17-3\sqrt{17}}} &0 \\
0 &0  &0 &1\\
\end{pmatrix},  
B_{4}=\begin{pmatrix}
1 &0 &0 &0\\
0 &1 &0 &0\\
0 &0 &\frac{\sqrt{2}}{2} &\frac{\sqrt{2}}{2}\\
0 &0 &\frac{\sqrt{2}}{2} &-\frac{\sqrt{2}}{2}
\end{pmatrix}, \label{beam22}
\end{align} are Bogoliubov transformations and can be realized using beam splitters.

\end{document}